\documentclass[aps,prl,reprint,superscriptaddress,nofootinbib,preprintnumbers]{revtex4-2}
\usepackage{graphicx}% Include figure files
\usepackage{dcolumn}% Align table columns on decimal point
\usepackage{bm}% bold math
\usepackage{braket}
\usepackage{epstopdf}

\usepackage{amsmath,amssymb,bm}
\usepackage{mathtools}
\usepackage{xcolor}
\usepackage{physics}
\usepackage{comment}

\usepackage{colortbl}

\newcommand{\Ra}{\rangle}
\newcommand{\La}{\langle}
\newcommand{\p}{\partial}

\begin{document}
%\arxivnumber{} % if you have one
\preprint{YITP-24-164}
\title{\boldmath Derivation of the GKP-Witten relation by symmetry without Lagrangian}

\author{Sinya Aoki}
\email{saoki@yukawa.kyoto-u.ac.jp} 
\affiliation{Center for Gravitational Physics and Quantum Information, Yukawa Institute for Theoretical Physics, Kyoto University, Kitashirakawa Oiwakecho, Sakyo-ku, Kyoto 606-8502, Japan}
\author{Janos Balog}
\email{balog.janos@wigner.hu} 
\affiliation{Holographic QFT Group, Institute for Particle and Nuclear Physics, Wigner Research Centre for Physics,  H-1525 Budapest 114, P.O.B. 49, Hungary}
\author{Kengo Shimada}
\email{kengo.shimada@yukawa.kyoto-u.ac.jp}
\affiliation{Center for Gravitational Physics and Quantum Information, Yukawa Institute for Theoretical Physics, Kyoto University, Kitashirakawa Oiwakecho, Sakyo-ku, Kyoto 606-8502, Japan}

\begin{abstract}
We derive the GKP-Witten relation in terms of correlation functions by symmetry without referring to a Lagrangian or the large $N$ expansion.
By constructing bulk operators from boundary operators in conformal field theory (CFT) by the conformal smearing, 
we first determine bulk-boundary 2-pt functions for an arbitrary spin using both conformal and bulk symmetries, then evaluate their
small $z$ behaviors, where $z$ is the $(d+1)$-th coordinate in the bulk.
Next, we explicitly determine small $z$ behaviors of bulk-boundary-boundary 3-pt functions also by the symmetries,
while small $z$ behaviors of correlation functions among one bulk and $n$ boundary operators with $n\ge 3$
are fixed by the operator product expansion (OPE).
Combining all results, we construct the GKP-Witten relation in terms of these correlation functions at all orders in an external source $J$.
We compare our non-Lagrangian approach with the standard approach employing the bulk action.  
Our results indicate that the GKP-Witten relation holds not only for holographic CFTs
but also for generic CFTs as long as certain conditions are satisfied.
\end{abstract}
%\pacs{04.20.-q, 04.20.Cv, 04.70.Dy}
\maketitle
\flushbottom

%%%%%%%%%%%%%%%%%%%%%%%%%%%%%%%%%%%%%%%

\section{Introduction}
%%%%%%%%%%%%%%%%%%%%%%%%%%%%%%%%%%%%%%%
The AdS/CFT correspondence~\cite{Maldacena:1997re,Gubser:1998bc,Witten:1998qj} states that observables of a ($d+1$)-dimensional
gravity theory on asymptotically anti-de Sitter (AdS) spacetime have dual description as operators in a $d$-dimensional CFT
living at the boundary of AdS. A partial explanation of the duality between these seemingly different physical models is the isomorphism between the
geometric isometry of the bulk AdS spacetime and the conformal symmetry group of the boundary CFT.
In the simple situation of a bulk scalar field ${\Phi}$ and a boundary (single trace) primary field $S$ (of conformal weight $\Delta_S$)
the dual relation is given by the BDHM \lq\lq dictionary'' \cite{Banks:1998dd} as
$ \lim_{z\to 0} z^{-\Delta_S}{\Phi}(z,x) =  S(x) $,
where taking the AdS boundary limit corresponds to $z\to 0$. 

An alternative, but fully equivalent formulation of the correspondence is the GKP-Witten formula conjectured by Gubser, Klebanov,
Polyakov \cite{Gubser:1998bc} and Witten \cite{Witten:1998qj}.
The general statement is that the bulk partition function %in the presence of sources coupled to the fields at the boundary is
with a particular boundary condition is equivalent to
the generating functional of the CFT correlation functions.
In the large $N$ limit where the problem can be treated semi-classically \cite{Skenderis:2002wp,Andrade:2006pg} and for a bulk scalar
field $\Phi$ with mass squared $M^2=\Delta_S(\Delta_S-d)$ in  AdS units, 
its classical solution in the presence of a source $J(x)$ for $\Phi$'s boundary dual $S(x)$ has small $z$ behavior
\begin{eqnarray}
     \Phi_{\rm cl}(z,x) = c z^{d-\Delta_S} J(x) + z^{\Delta_S} \phi(x) +\dots
\label{int1}
\end{eqnarray}
where the dots stand for subleading terms in $z$ and
\begin{eqnarray}
     \phi(x) = \langle S(x)\rangle_J=\langle S(x)\exp\big\{\!\int \! {\rm d}^d yS(y)J(y)\big\}\rangle_c
\label{int2}
\end{eqnarray}
is the CFT expectation value (1-point function) of $S(x)$ in the presence of the source.
The 1-point function can be represented as
\begin{eqnarray}
&&\La S(x)\Ra_J =\int {\rm d}^d y {J(y)\over \vert x-y\vert^{2\Delta_S}} + \label{int3} \\
&&\frac{1}{2}\int{\rm d}^dy_1{\rm d}^dy_2\La S(x)S(y_1)S(y_2)\Ra_{c}\, J(y_1)J(y_2)+
{\rm O}(J^3).  \nonumber
\end{eqnarray}
Here we used the explicit form of the (normalized) 2-point function $\La S(x)S(y)\Ra$.

Even after several decades of intense studies, the emergence of the extra dimension remains somewhat mysterious.
It is generally believed \cite{Heemskerk:2010hk} that the emerging AdS radial direction has to do with the energy scale of a kind of 
renormalization group transformation. 
One possibility is to use the multi-scale entanglement renormalization ansatz
(cMERA) \cite{Nozaki:2012zj,Miyaji:2015yva,Miyaji:2015fia},
where the extra dimension is interpreted as the level of the coarse-graining. 
Following the same philosophy, a similar but different coarse-graining technique was proposed by some of the present
authors~\cite{Aoki:2015dla,Aoki:2016ohw,Aoki:2017bru,Aoki:2017uce,Aoki:2018dmc}. In this approach, the a priori existence of an
AdS spacetime is not assumed, and the bulk spacetime is constructed from the CFT in such a way that the metric can be interpreted as an
information metric associated with a state of the boundary~\cite{Aoki:2017bru}.

Recently an improved version of this approach was proposed~\cite{Aoki:2022lye}, which can be called the conformal smearing 
since in this case, the conformal transformations of the boundary CFT fields induce bulk coordinate transformations on the bulk fields, which 
become isometries of the would-be AdS.
In~\cite{Aoki:2022lye}, the GKP-Witten relation (for a scalar field) was verified up to order $J$.

In this paper, we extend our study of the GKP-Witten relation in terms of correlation functions without using a Lagrangian
to an arbitrary spin at all orders in $J$. 
Employing the embedding formula~\cite{Costa:2011mg,Costa:2014kfa} to deal with spins, we evaluate bulk-boundary 2-pt functions and
bulk-boundary-boundary 3-pt functions. Beyond these cases, the small $z$ behaviors of higher-point functions are directly evaluated
by the OPE and mathematical induction.
We use these results to derive the GKP-Witten relation for an arbitrary spin and all orders in $J$. We compare our approach with the
standard derivation using the bulk action. Our derivation implies
that the GKP-Witten relation holds for a generic CFT thanks to the symmetry considered in this paper.

%%%%%%%%%%%%%%%%%%%%%%%%%%%%%%%%%%%%%%%

\section{Conformal smearing approach}
We start with the $O(N)$ invariant scalar conformal field theory in $d$ dimensional Euclidean space, whose non-singlet primary field satisfies
\begin{eqnarray}
\langle 0 \vert \hat\varphi^a(x) \hat\varphi^b(y)\vert 0 \rangle = \frac{\delta^{ab}}{N} \frac{1}{\vert x-y\vert^{2\Delta_\varphi}}, 
\end{eqnarray}
where $a,b=1,2,\cdots, N$ and $\Delta_\varphi$ is the conformal dimension of $\hat\varphi$. In the case of  $N=1$, the theory is a $Z_2$ invariant CFT.  
We require $\Delta_\varphi< d/2$ in our approach.

We define a bulk field on $d+1$ dimensional space by the conformal smearing~\cite{Aoki:2022lye} as
\begin{eqnarray}
\hat\sigma^a(X) = \int d^d y \, h(z,x-y) \hat\varphi^a(y),  %\quad  X=(z,x)
\end{eqnarray}
where $X^0 = z,~X^\mu = x^\mu$, $\mu \in ( 1,\cdots,d)$ and the smearing kernel is given by
\begin{eqnarray}
&&h(z,x) = \Sigma_0 \left(\frac{z}{x^2+z^2}\right)^{d-\Delta_\varphi}, \\
&&\Sigma_0 :=\sqrt{{\Gamma(d)\Gamma(d-\Delta_\varphi)\over N \pi^d \Gamma({d\over 2} -\Delta_{\varphi})\Gamma({d\over 2})}}, \nonumber
\end{eqnarray}
so that $\langle 0\vert \hat\sigma^a(X)\sigma^b(X)\vert0\rangle =\delta^{ab}/N$.
This form of the smearing kernel agrees with the one in the HKLL bulk reconstruction, which is however defined for a singlet operator $\Phi$ with $\Delta_\Phi > d-1$~\cite{Hamilton:2006az}.
 
The conformal smearing maps the conformal transformation on $\hat\varphi(x)$ to a part of the general coordinate transformation on $\hat\sigma^a(X)$, which agrees with an AdS isometry~\cite{Aoki:2022lye}.
Explicitly the unitary operator $U$ for the conformal transformation is defined by
\begin{equation}
    U\hat\varphi^a(x) U^\dagger = h(x)^{\Delta_\varphi}\hat\varphi^a(\tilde x),
\end{equation}
where (1) Poincare transformation: $\tilde x^\mu = \Omega^\mu{}_\nu x^\nu+a^\mu $, $h(x)=1$, (2) dilatation: $\tilde x^\mu= \lambda x^\mu$, $h(x)=\lambda$ and (3)
inversion\footnote{
The special conformal transformation is constructed by an inversion followed by translation and inversion again.}: $\tilde x^\mu=x^\mu/x^2$, $h(x)=1/x^2$,
generates the bulk coordinate transformation as
\begin{equation}
    U\hat\sigma^a(X) U^\dagger = \hat\sigma^a(\tilde X), \label{unitary_bulk}
\end{equation}
where (1) $\tilde x^\mu = \Omega^\mu{}_\nu x^\nu + a^\mu$ and $\tilde z=z$, (2) $\tilde X^M =\lambda X^M$ and (3) $\tilde X^M= X^M/(x^2 + z^2)$ with $M \in (0,1,\cdots,d)$.  

In this paper, we consider bulk-boundary correlation functions given by
\begin{eqnarray}
&&G^{1,p_2,\cdots,p_n}_{A,s_2,\cdots,s_n}(X_1,x_2,\cdots,x_n)  \label{eq:correation_gen} \\
&&\hspace{4em}:= \langle 0\vert G^1_{A}(X_1)O^{p_2}_{s_2}(x_2) \cdots O^{p_n}_{s_n}(x_n) \vert 0 \rangle,
\nonumber
\end{eqnarray}
where $G^1_{A}$ is an $O(N)$ invariant spin $L:=\vert A\vert$ operator in the bulk constructed from $\hat\sigma^a$'s and $\p_{M} \hat\sigma^a$'s,
and $A:=M^1\cdots M^L$ is a compact notation for its spin $L$ index (symmetric traceless as a bulk tensor)
and $\vert A\vert$ is the length of $A$,
while $O^{p}_{s}$ is the spin $L_{p}=\vert s\vert$ conformal primary
with $s:=\mu^1\cdots \mu^{L_p}$%, $\mu \in ( 1,\cdots,d)$
and conformal dimension $\Delta_p$. In this paper, we use implicit notations that all operators are ``time'' ordered in terms of the radial coordinate
$\vert x\vert$: $O(x) O'(x')$ means $O(x) O'(x') $ for $\vert x\vert> \vert x'\vert$ but $O'(x') O(x) $ for $\vert x'\vert> \vert x\vert$. 
These correlation functions must satisfy 
\begin{eqnarray}
&&G^{1,p_2,\cdots,p_n}_{\tilde{A},\tilde{s}_2,\cdots,\tilde{s}_n} (\tilde X_1,\tilde x_2,\cdots,\tilde x_n) \nonumber \\
&&\hspace{1em}= G^{1,p_2,\cdots,p_n}_{A,s_2,\cdots,s_n} ( X_1, x_2,\cdots, x_n) \times
\label{eq:constraint} \\
&& \hspace{3em}   {\partial X_1^{A}\over  \partial \tilde X_1^{\tilde{A}}}
h(x_2)^{-\Delta_{p_2}}
    {\partial x_2^{s_2}\over  \partial \tilde x_2^{\tilde{s}_2}} \cdots h(x_n)^{-\Delta_{p_n}} {\partial x_n^{s_n}\over  \partial \tilde x_n^{\tilde{s}_n}}   , \nonumber
\end{eqnarray}
where the short-hand notation means
\begin{eqnarray}
 {\partial X^{A}\over  \partial \tilde{X}^{\tilde{A}}}:= \prod_{k=1}^L {\partial X^{M^k}\over  \partial \tilde{X}^{\tilde{M}^k}} , \qquad
  {\partial x^{s}\over  \partial \tilde x^{\tilde{s}}}:= \prod_{k=1}^L {\partial x^{\mu^k}\over  \partial \tilde x^{\tilde{\mu}^k}} .
\end{eqnarray}
While the conformal smearing is necessary to guarantee that such $G^{1}_{A}(X)$'s indeed exist,
we will not use its detailed properties except the constraint \eqref{eq:constraint} in the rest of this paper.

\section{Derivation of the GKP-Witten relation for an arbitrary spin}
In this section, we derive the GKP-Witten relation for spinning fields at all order in $J$, using the constraint \eqref{eq:constraint}.

\subsection{2-pt function}
The bulk-boundary 2-pt function $G^{1,p_2}_{A_1,s_2}(X_1,x_2):=\langle 0\vert G^{1}_{A_1}(X_1)O^{p_2}_{s_2}(x_2) \vert 0 \rangle$ can be easily obtained in the embedding space as~\cite{Costa:2011mg,Costa:2014kfa},
\begin{eqnarray}
    &&G(X_1,W_1;P_2,Q_2)  \\
    &&= c {\left[ (-2X_1\cdot P_2) (W_1\cdot Q_2) +(2W_1\cdot P_2) (X_1\cdot Q_2)\right]^{L_{p_2}}\over (-2 X_1\cdot P_2)^{\Delta_{p_2} + L_{p_2}}}, \nonumber
\end{eqnarray}
where $c$ is a constant, $X_1=(1/z_1,(z_1^2+x_1^2)/z_1,x^\mu/z_1)$ and $P_2=(1,x_2^2,x_2^\mu)$ in the light-cone coordinate in $\mathbb{R}^{d+1,1}$, while  $W_1$ and $Q_2$ take care of spin indices $A_1$ and $s_2$, respectively.
Note that the bulk-boundary 2-pt function can be determined by either the constraint \eqref{eq:constraint} or the bulk free equations of motion (EOM). 

Since $-2X_1\cdot P_2= (z_1^2 +(x_1-x_2)^2)/z_1$ and $W_1=Q_1/z_1$, we obtain
\begin{eqnarray}
    &&\lim_{z_1\to 0} G(X_1,W_1;P_2,Q_2) = c z_1^{-L_{p_2}} \times \\
    &&  \left[ z_1^{\Delta_{p_2}} G(P_1,Q_1; P_2,Q_2) + z_1^{d-\Delta_{p_2}} {(Q_1\cdot Q_2)^{L_{p_2}}\over \Lambda_{\Delta_{p_2}}} \delta^{(d)}(x_{12}) \right] \nonumber
\end{eqnarray}
for $\Delta_{p_2}>d/2$ (otherwise the 2nd term is absent), 
where $G(P_1,Q_1; P_2,Q_2)$ corresponds to the 2-pt function in CFT, and we use
\begin{eqnarray}
\hspace{-2em}\lim_{z\to 0} {1\over (z^2+x_{12}^2)^\alpha} = (-2 P_1\cdot P_2)^{-\alpha}+{z^{d-2\alpha}\over \Lambda_\alpha}\delta^{(d)}(x_{12}) 
\end{eqnarray}
with $1/ \Lambda_\alpha =\int d^dx  (x^2+1)^{-\alpha}$.
By restricting $A_1$ to $s_1$, symmetric and traceless in $d$ dimensions, this reads 
\begin{eqnarray}
\lim_{z_1\to 0}  G^{1,p_2}_{s_1,s_2}(X_1,x_2) &=& c z_1^{\Delta_{p_2}-L_{p_2}} \langle 0\vert O^{p_2}_{s_1}(x_1)O^{p_2}_{s_2}(x_2) \vert 0 \rangle \nonumber \\
&+& 
{c z_1^{d-\Delta_{p_2}-L_{p_2}}\over \Lambda_{\Delta_{p_2}}} \delta_{s_1,s_2} \delta^{(d)}(x_{12}) \nonumber\\
&=:& c F^{1,p_2}_{s_1,s_2}(X_1,x_2) \label{eq:2-pt_limit}
\end{eqnarray}
in the standard notation, where the last term is absent if $\Delta_{p_2}\le d/2$.

\subsection{Diagonalization of bulk operators}
We assume that there are $k$ conformal primary operators with spin $L_{p_1}=L_{p_2}=\cdots=L_{p_k}=\vert s\vert$ denoted by $O^{p_1}_{s}, O^{p_2}_{s},\cdots O^{p_k}_{s}$
whose conformal dimensions $\Delta_{p_i}$ satisfy
$\Delta_{p_1} < \Delta_{p_2}<\cdots < \Delta_{p_k}$
without degeneracy.

In the standard approach, we assume from the beginning that the bulk operator $\Phi^i_{s}$ couples only to one specific primary operator $O^{p_i}_s$
since $\Phi^i_{s}$ satisfies the bulk free EOM with a specific mass determined by $\Delta_{p_i}$.
Our bulk operator $G^1_{s}$, however, couples to all $O^{p_i}_{s}$ ($i=1,2,\cdots,k$) in general. Since bulk operators are composite operators of $\hat\sigma^a$
determined from $\hat\varphi^a$ by the conformal smearing, there must be
$k$ independent bulk operators $\bar{G}^{i}_{s}$  ($i=1,2,\cdots,k$) which satisfy
\begin{eqnarray}
\hspace{-1em}\lim_{z_1\to 0}  \langle 0\vert \bar{G}^{i}_{s_1}(X_1) O^{p_j}_{s_2}(x_2)\vert 0\rangle 
 = c_{ij} F^{i,p_j}_{s_1,s_2}(X_1,x_2),
\end{eqnarray}
where the "independent" means that $c_{ij}$ has an inverse $c^{-1}_{ij}$. 
We then construct a set of  bulk operators as $G^{i}_{s}=\sum_j c_{ij}^{-1} \bar{G}^{j}_{s}$, which is diagonal as
\begin{eqnarray}
 \hspace{-1em} \lim_{z_1\to 0}  \langle 0\vert G^{i}_{s_1}(X_1) O^{p_j}_{s_2}(x_2)\vert 0\rangle 
 = \delta_{ij} F^{i,p_i}_{s_1,s_2}(X_1,x_2).
\end{eqnarray}
Thus $G^{i}_{s}$ couples only to $O^{p_i}_{s}$ and the corresponding 2-pt function $ \langle 0\vert G^{i}_{s_1}(X_1) O^{p_i}_{s_2}(x_2)\vert 0\rangle$ satisfies
the EOM for the spin $L_{p_i}$ particle in AdS with mass squared $M^2=\Delta_{p_i}(\Delta_{p_i}-d)-L_{p_i}$ in AdS units~\cite{Costa:2014kfa}.

\subsection{3-pt functions}
We next investigate the 3-pt function.
Using the embedding formula~\cite{Costa:2014kfa}, the bulk-boundary-boundary 3-pt function is given in the standard notation as
\begin{eqnarray}
&&\langle 0\vert G^{i}_{s_1}(X_1) O^{p_2}_{s_2}(x_2) O^{p_3}_{s_3}(x_3) \vert 0\rangle   =    H_{s_1s_2s_3}(X_1,x_2,x_3) 
\nonumber\\
&&\times \left({z_1\over z_1^2+(x_1-x_2)^2}\right)^{\Delta_{p_2}+L_{p_2}}
\! \! \! \left({z_1\over z_1^2+(x_1-x_3)^2}\right)^{\Delta_{p_3}+L_{p_3}}
\nonumber \\
&&~~~~ \times  F\left( {z_1^2 (x_2-x_3)^2\over
[z_1^2+(x_1-x_2)^2] [z_1^2+(x_1-x_3)^2]}\right),  \label{eq:3pt_Bbb}
\end{eqnarray}
where $F$ is an arbitrary function, and $H_{s_1s_2s_3}$ takes care of the spin indices and $\lim_{z_1\to 0}H_{s_1s_2s_3} = z_1^{-2L_{p_i}} h_{s_1s_2s_3}$ with $h_{s_1s_2s_3}$ being the
spin factor in CFT.

On the other hand, we can use the OPE given by
\begin{eqnarray}
\hspace{-2em}O^{p_2}_{s_2}(x_2) O^{p_3}_{s_3}(x_3)
    = \sum_{p,s} C^{p_2p_3p}_{s_2s_3s}\left(x_2-x_3, \partial^{x_3}\right) O^{p}_{s}(x_3),
\end{eqnarray}
where $p$ runs over all primary operators and derivatives $\p^{x_3}$ generate their descendants. Since both $G^{i}_{s_1}$ and $O^{p_i}_{s_1}$ couple to only the $p=p_i$ terms in the OPE expansion,
we obtain, for $\vert x_1\vert > \vert x_2\vert, \vert x_3\vert$ or  $\vert x_1\vert < \vert x_2\vert, \vert x_3\vert$, 
\begin{eqnarray}
&&\lim_{z_1\to 0} \langle 0\vert G^{i}_{s_1}(X_1) O^{p_2}_{s_2}(x_2) O^{p_3}_{s_3}(x_3) \vert 0\rangle  \label{eq:3pt_bbb} \\
&&= z_1^{\Delta_{p_i}-L_{p_i}} \langle 0\vert  O^{p_i}_{s_1}(x_1) O^{p_2}_{s_2}(x_2) O^{p_3}_{s_3}(x_3)\vert 0\rangle +\cdots \nonumber
\end{eqnarray}
where $+\cdots$ represents possible contributions from the delta function in eq.~\eqref{eq:2-pt_limit}, which are usually subtracted by boundary
counter terms in the standard semiclassical approach~\cite{Skenderis:2002wp,Andrade:2006pg} to the GKP-Witten relation.
We therefore neglect these types of contributions hereafter except those in the 2-pt function. 
Since eq.~\eqref{eq:3pt_Bbb} should agree with the above expression, we see that $F(x)\propto x^\alpha$ as $x\to 0$, where
$2\alpha=\tau_{p_2}+\tau_{p_3}-\tau_{p_i}$ with $\tau_{p}=\Delta_{p}+L_{p}$.
Therefore eq.~\eqref{eq:3pt_bbb} holds for any operator orderings.

\subsection{Bulk-boundary $n$-pt functions}
We consider the Bulk-boundary correlation functions, defined by
\begin{eqnarray}  &&G^{i,p_2,\cdots,p_n}_{s_1,s_2,\cdots,s_n}(X_1,x_2,\cdots, x_n) \\
&&:= \langle 0\vert G^{i}_{s_1}(X_1) O^{p_2}_{s_2}(x_2)\cdots O^{p_n}_{s_n}(x_n)\vert 0\rangle \nonumber
\end{eqnarray}
for $n\ge 3$.
We will show that
\begin{eqnarray}
&&\lim_{z_1\to 0} G^{i,p_2,\cdots,p_n}_{s_1,s_2,\cdots,s_n}(X_1,x_2,\cdots, x_n) \label{eq:n-pt_OPE} \\
&&= z_1^{\Delta_{p_i}-L_{p_i}}\langle 0\vert O^{p_i}_{s_1}(x_1) O^{p_2}_{s_2}(x_2)\cdots O^{p_n}_{s_n}(x_n)\vert 0\rangle~~~~\nonumber
\end{eqnarray}
by the mathematical induction.
In the previous subsection, we have shown eq.~\eqref{eq:n-pt_OPE} for $n=3$.
We assume that eq.~\eqref{eq:n-pt_OPE} holds for $n$. 
Using the OPE, we have
\begin{eqnarray}
&&G^{i,p_2,\cdots,p_{n+1}}_{s_1,s_2,\cdots,s_{n+1}}(X_1,x_2,\cdots,x_n, x_{n+1}) \nonumber\\
&& =\sum_{p,s} C^{p_{n}p_{n+1} p}_{s_{n} s_{n+1}s}\left(x_n-x_{n+1}, \partial^{x_{n+1}}\right) \\
&& ~~~~~ \times  G^{i,p_2,\cdots,p_{n-1},p}_{s_1,s_2,\cdots,s_{n-1},s}( X_1,x_2,\cdots, x_{n-1},x_{n+1}) . \nonumber
\end{eqnarray}
Therefore eq.~\eqref{eq:n-pt_OPE} holds for $n+1$ as
\begin{eqnarray}
&&\lim_{z_1\to 0} G^{i,p_2,\cdots,p_{n+1}}_{s_1,s_2,\cdots,s_{n+1}} (X_1,x_2,\cdots, x_n,x_{n+1}) \\
&&=  z_1^{\Delta_{p_i}-L_{p_i}}\sum_{p,s} C^{p_{n}p_{n+1} p}_{s_{n} s_{n+1}s}\left(x_n-x_{n+1}, \partial^{x_{n+1}}\right)\nonumber \\
&& \hspace{1em} \times    
\langle 0\vert O^{p_i}_{s_1}(x_1) O^{p_2}_{s_2}(x_2)\cdots O^{p_{n-1}}_{s_{n-1}}(x_{n-1}) O^{p}_{s}(x_{n+1})\vert 0\rangle  \nonumber \\
&&=  z_1^{\Delta_{p_i}-L_{p_i}} 
\langle 0\vert O^{p_i}_{s_1}(x_1) O^{p_2}_{s_2}(x_2)\cdots O^{p_{n+1}}_{s_{n+1}}(x_{n+1})\vert 0\rangle. \nonumber
\end{eqnarray}
Thus the formula eq.~\eqref{eq:n-pt_OPE} is proven by the mathematical induction.

\subsection{GKP-Witten relation}
In our non-Lagrangian approach, the GKP-Witten relation is defined in terms of correlation functions.
We define, for a bulk spin $L_{p_i}=\vert s_1\vert$ operator in the presence of sources in CFT, the analog of $\Phi_{\rm cl}$ as
\begin{eqnarray}
\Phi^{i}_{s_1}(X_1) &:=&\frac{
   \langle 0\vert G^{i}_{s_1}(X_1)
    \exp\left[
    \int {\rm d}^dy\,  \sum_{q,s} J^{q,s}(y) O^{q}_{s}(y)\right]\vert 0\rangle}{Z(J)} \nonumber\\
    &:=&\sum_{n=1}^\infty  \Phi_{s_1}^{i(n)}(X_1),
    \label{eq:all}
\end{eqnarray}
where $\Phi_{s_1}^{i(n)}$ contains ${\rm O}(J^n)$ terms, and $Z(J)$, defined similarly as the numerator without $G^i_{s_1}$,
removes disconnected contributions.

For $n=1$, it is easy to see that
\begin{eqnarray}
&&\lim_{z_1\to 0}  \Phi_{s_1}^{i(1)}(X_1) =  {z_1^{d-\Delta_{p_i}-L_{p_i}}\over \Lambda_{\Delta_{p_i}}}J^{p_i,s_1}(x_1) + \\
&&\hspace{1em} z_1^{\Delta_{p_i}-L_{p_i}}\int {\rm d}^dx_2\,\sum_s J^{p_i,s}(x_2) \langle 0\vert O^{p_i}_{s_1}(x_1)O^{p_i}_{s}(x_2)\vert 0\rangle_c , \nonumber \\
%&& \hspace{1em}+ {z_1^{d-\Delta_{p_i}-L_{p_i}}\over \Lambda_{\Delta_{p_i}}}J^{p_i,s_1}(x_1) , \nonumber
\end{eqnarray}
while, for $ n-1 \ge 2$, we obtain 
\begin{eqnarray}
&&\lim_{z_1\to 0}  \Phi_{s_1}^{i(n-1)}(X_1) \\
&&=\frac{z_1^{\Delta_{p_i}-L_{p_i}}}{(n-1)!}
\int {\rm d}^dx_2\cdots {\rm d}^dx_{n}\, J^{p_2,s_2}(x_2)\cdots  J^{p_n,s_n}(x_{n}) \nonumber \\
&&\hspace{10em}\times \langle 0\vert O^{p_i}_{s_1}(x_1) O^{p_2}_{s_2}(x_2)\cdots O^{p_n}_{s_n}\vert 0\rangle_c, \nonumber 
\end{eqnarray}
where we neglect contributions from delta functions, as mentioned before.

Employing the above formula, we can establish the GKP-Witten relation, which is completely analogous to (\ref{int1}) and (\ref{int2}), as
\begin{eqnarray}
&&\lim_{z_1\to 0} \Phi^{i}_{s_1}(X_1) = {z_1^{d-\Delta_{p_i}-L_{p_i}}\over \Lambda_{\Delta_{p_i}}}J^{p_i,s_1}(x_1) + \cdots 
\label{GKPWres}\\
&+&z_1^{\Delta_{p_i}-L_{p_i}} \langle 0\vert O^{p_i}_{s_1}(x_1)
  \exp\left[ \int {\rm d}^dy\,  \sum_{p,s} J^{p,s}(y) O^{p}_{s}(y)\right]\vert 0\rangle_c \nonumber  
%  &&\hspace{1em}+ {z_1^{d-\Delta_{p_i}-L_{p_i}}\over \Lambda_{\Delta_{p_i}}}J^{p_i,s_1}(x_1) +\cdots \label{GKPWres}
\end{eqnarray}
for $i=1,2,\cdots, k$, where $\cdots$ represent local contributions,
which contain  $J(x_1)$, and are usually subtracted by boundary counter terms in the standard Lagrangian approach.   

\subsection{Some comparisons with the standard calculation} 
Here, we make some comparisons between the standard method based on bulk action and our approach based on symmetries.

In the standard approach, $n$-pt functions in CFT with $n\ge 3$ are calculated from Witten diagrams in the bulk.
The 3-pt function, for example, reads~\cite{Giombi:2009wh,Giombi:2010vg,Costa:2014kfa}
\begin{eqnarray}
  &&  \langle 0\vert O^{p_1}_{s_1}(x_1) O^{p_2}_{s_2}(x_2) O^{p_3}_{s_3}(x_3) \vert 0\rangle_{\rm CFT} \, 
 \stackrel{?}{=} \int d^{d+1}X\, g^{u_1u_2u_3}_{p_1p_2p_3}(X) \nonumber \\
&&\times G^{p_1}_{u_1 s_1}(X,x_1) \label{eq:GKPW_full} G^{p_2}_{u_2 s_2}(X,x_2)  G^{p_3}_{u_3 s_3}(X,x_3)\\
&&= \langle 0\vert O^{p_1}_{s_1}(x_1) O^{p_2}_{s_2}(x_2) O^{p_3}_{s_3}(x_3) \nonumber \\
&&\hspace{1em}\times \int d^{d+1}X\, g^{u_1u_2u_3}_{q_1q_2q_3}(X)
    \Phi^{q_1}_{u_1}(X) \Phi^{q_2}_{u_2}(X) \Phi^{q_3}_{u_3}(X) \vert 0\rangle,
\end{eqnarray}
where $g^{u_1u_2u_3}_{p_1p_2p_3}(X)$ is the bulk 3-pt vertex function and $G^{p}_{us}(X,y)$ is the bulk-boundary propagator.
It is rather involved to confirm the first equality explicitly.
On the other hand, since the bulk 3-pt vertex with bulk operators after the $X$ integration transforms trivially under the bulk transformation,
the expression after the second equality has the same transformation property as the 3-pt function in CFT.
This fact establishes the first equality without any calculations up to normalization, 
{when 3-pt functions in CFT are uniquely given, for example, in the case of the scalar 3-pt function.
If 3-pt functions in CFT have several tensor structures for spins\cite{Costa:2011mg}, however, explicit calculations ane necessary to confirm the first equality\cite{Giombi:2009wh,Giombi:2010vg,Costa:2014kfa}.\footnote{We would like to thank Prof.~Yu Nakayama for his useful comment. } 

The $n$-pt function for $n\ge 4$ constructed in a similar manner also has the same transformation property as the $n$-pt function in CFT, which however does not guarantee that they agree due to the non-uniqueness of the $n$-pt function in CFT. 
Explicit calculations are necessary.

\section{Conclusion}

In this paper we derive the GKP-Witten relation, eq.~(\ref{GKPWres}), in terms of correlation functions by using only symmetries without a Lagrangian.
Eq.~(\ref{GKPWres}) holds for all spin and primary operators in the presence of arbitrary sources coupled to the primaries.
Constraints \eqref{eq:constraint} together with the OPE are sufficient to prove eq.~(\ref{GKPWres}),
while the conformal smearing is only used to derive \eqref{eq:constraint}. 

Our results indicate that 
the GKP-Witten relation holds not only for holographic CFTs in the standard semi-classical analysis in the large $N$ expansion but also for an arbitrary
CFT without the large $N$ expansion.
This means that the GKP-Witten relation cannot be used to judge whether 
an AdS/CFT duality comes from a holographic CFT or not.
While the bulk-boundary 2-pt function satisfies the free EOM in the bulk AdS, connected higher point functions are non-zero at the same time,
indicating that the bulk theory is interacting. 

\vskip 0.4cm

In this paper, we only considered correlation functions involving one bulk operator and $n$ boundary operators, which is enough to show
eq.~(\ref{GKPWres}). It is interesting and important to extend our analysis to correlation functions involving more than one bulk operator|SA{s}.
It is also important to estimate effects coming from terms in the boundary theory  which break conformal symmetry such as
$\lambda \varphi^4$ in 3-dimensions~\cite{Aoki:2024opu}.

Our results may provide a new approach to the AdS/CFT correspondence, where
the bulk theory is defined in terms of correlation functions without Lagrangian/action and even the CFT Lagrangian is not required.
Therefore one may investigate higher spin theories with our approach,
as it is expected to be dual to the $O(N)$ vector model. 
An extension of the conformal bootstrap method~\cite{Ferrara:1973yt,Polyakov:1974gs,Mack:1975jr} to  bulk-boundary correlation functions would also be interesting.

%%%%%%%%%%%%%%%%%%%%%%%%%%%%%%%%%%%%%%%

\section*{Acknowledgment}
J.B. acknowledges financial support from the International Research Unit of Quantum
Information (QIU) of Kyoto University Research Coordination Alliance, 
and kind hospitality during his stay at Yukawa Institute for Theoretical Physics, Kyoto University. 
This work has been supported in part by the JSPS Grant-in-Aid for Scientific Research (No. JP22H00129) and the NKFIH grant K134946.

%\appendix

%\bibliographystyle{apsrev4-2}
\bibliographystyle{utphys}
\bibliography{BulkReconstruction}

%%%%%%%%%%%%%%%%%%%%%

\end{document}